\documentclass{article}
\usepackage{ltwol2e}
\usepackage{epsfig}
\usepackage{rotating}

\def\e       { {\mathrm{e}} }
\def\E       { {\mathrm{E}} }
\def\W       { {\mathrm{W}} }
\def\Z       { {\mathrm{Z}} }

\def\enw     { \e \nu \W }
\def\nng     { \nu \bar{\nu} \gamma }

\def\eeZ     { \e \e \Z }
\def\stat    { {\mathrm{stat.}} }
\def\syst    { {\mathrm{syst.}} }

\def\M       { {\mathrm{M}} } 
\def\MeV     { {\mathrm{MeV}} }

\def\GeV     { {\mathrm{GeV}} }

\def\GeVcc   { \GeV/c^2 }
\def\pb      { {\mathrm{pb}} }

\def\WWZ     { {\mathrm{WWZ}} }
\def\WWg     { {\mathrm{WW}\gamma} }
\def\WW      { {\mathrm{WW}} }
\def\Wg      { {\mathrm{W}\gamma} }
\def\gg      { g_{1}^{\gamma} }
\def\gZ      { g_{1}^{\mathrm{Z}} }
\def\kg      { \kappa_{\gamma} }
\def\kZ      { \kappa_{\mathrm{Z}} }
\def\dkg     { \Delta\kappa_{\gamma} }
\def\dkZ     { \Delta\kappa_{\mathrm{Z}} }
\def\dgZ     { \Delta\gZ }
\def\lg      { \lambda_{\gamma} }
\def\lZ      { \lambda_{\mathrm{Z}} }
\def\qq      { {\mathrm{q}}{\bar{\mathrm{q}}'}}

\def\dM      { \Delta{\mathrm{M}}}

\def\Journal#1#2#3#4{{#1} {\bf #2}, #3 (#4)}

\def\CPC{\em Comp. Phys. Comm.}
\def\IJMPA{{\em Int. Journal of Mod. Phys.} A}

\def\NPB{{\em Nucl. Phys.} B}
\def\PLB{{\em Phys. Lett.}  B}
\def\PRL{\em Phys. Rev. Lett.}
\def\PR{\em Phys. Rev.}
\def\PRD{{\em Phys. Rev.} D}
\def\ZPC{{\em Z. Phys.} C}

\begin{document}

\pagestyle{empty}

\parbox{16.0cm}{
\flushright{{\Huge \bf ECOLE}} \\[50pt] 
\flushright{{\Huge \bf POLYTECHNIQUE}} \\[50pt] 
\flushright{{\Huge \bf IN2P3-CNRS}} }

\vspace{3.5cm}

\hspace*{3.0cm}\parbox{11.0cm}{   

\flushright{{\large\bf X-LPNHE/98-08}} \\[50pt]
 
\begin{center}
 {\Large\bf Single W Production at LEP2} \\[30pt]
 {\large R.~Tanaka} \\[10pt]
 {\it LPNHE, Ecole Polytechnique, F-91128 Palaiseau CEDEX, FRANCE} \\[20pt]
 {\normalsize  Talk given at the XXIX International Conference} \\
 {\normalsize  on High Energy Physics, July 23-29, 1998, 
                Vancouver, CANADA.} \\[20pt] 
\end{center} }

\newpage

\title{SINGLE W PRODUCTION AT LEP2}

\author{R.~TANAKA}

\address{LPNHE, Ecole Polytechnique,
        Route de Saclay, F-91128 Palaiseau CEDEX, FRANCE\\
        E-mail: Reisaburo.Tanaka@in2p3.fr}   

\twocolumn[\maketitle\abstracts{ 
Single W and single gamma productions which are sensitive to the 
trilinear gauge coupling $\WWg$ have been studied at LEP.
It is shown that single W production has particular sensitivity to
the `anomalous' magnetic moment $\kg$ of the W boson,
complementary to WW production at LEP and $\Wg$ production at hadron
colliders.
The invisible decay of W boson has been searched and the limit on the 
invisible decay width of $27\:\MeV$ at $95\,\%\:$C.L. has been obtained. }]

\section{Introduction}

\pagestyle{plain}
\setcounter{page}{1}

   The existence of the trilinear gauge couplings (TGC) is the direct 
consequence of the non-Abelian 
$\mbox{SU}(2) \times \mbox{U}(1)$ 
gauge theory which has not been studied in detail.   
Precise measurements of these couplings make it possible to test the 
standard model.   Any deviation from the standard model would indicate
the new physics.   There are $2 \times 7$ parameters of couplings in the
effective Lagrangian~\cite{ref:TGC}.   By requiring C- and P-invariance,
also $\gg=1$ by electromagnetic gauge invariance, the number of
parameters reduces to 5:
$\dgZ \equiv \gZ-1,\ \dkg \equiv \kg-1,\ \dkZ \equiv \kZ -1,\ \lg$ 
and $\lZ$ where all these parameters are vanishing in the standard model.
For $\W^+$ boson, these parameters can be related to 
the electromagnetic charge: 
   $e_\W   = e\,\gg$, 
and the static moments~\cite{ref:moments} as,
magnetic dipole moment: 
   $\mu_\W = \frac{e}{2m_\W} (\gg + \kg + \lg)$,
and electric quadrupole moment: 
    $Q_\W   = -\frac{e}{m^2_\W} (\kg - \lg)$, 
and also those associated with the weak boson Z.

At LEP2, it became, for the first time in the $\e^+\e^-$ collider, 
to perform the direct measurement of TGC.   W pair production plays
a principal role to study $\WWg$ and $\WWZ$ couplings~\cite{ref:Yellow}.
However, these two couplings cannot be separated each other.
Single W production, 
$\e^+\e^- \rightarrow \enw$~\cite{ref:single_W,ref:Kurihara}, 
or single gamma production, 
$\e^+\e^- \rightarrow \nu \nu \gamma$~\cite{ref:single_gamma},
can be used to test the $\WWg$ coupling.  
At hadron colliders, $\W\gamma$ production has been studied to probe 
the $\WWg$ vertex~\cite{ref:UA2,ref:CDF,ref:D0}
where the form factor $\Lambda$ 
is introduced to assure the unitarity~\cite{ref:Baur}.   The TGC limits
derived at LEP are insensitive to the form factor scale and power.
To relate $\WWZ$ and $WWg$ couplings, SU(2)$\times$U(1)
constraints, $\dgZ = \dkZ + \dkg \tan^2\theta_\W$ and $\lZ = \lg$,
are imposed.

   In the search for supersymmetric particles, chargino pair production 
($\e^+\e^- \rightarrow \chi^+ \chi^-$) where charginos decay
predominantly into sneutrinos and leptons, 
it is experimentally difficult if the mass difference between chargino
and sneutrino is small.   This is due to the huge backgrounds from two
photon process.   Therefore the search for single W events in 
$\e^+\e^- \rightarrow \W^+ \W^-$ process where one W boson decays to 
undetected chargino and neutralino and the other W to the standard model
particles has been proposed~\cite{ref:susy}.
A search for this scenario has been performed by ALEPH.

All results presented in this paper are preliminary except L3 results.

\section{$\enw$ Production}

\subsection{Sensitivity to TGC\,($\WWg$)} 

\begin{figure}[b]
   \begin{center}
        \vspace{-1cm}
      \epsfig{figure=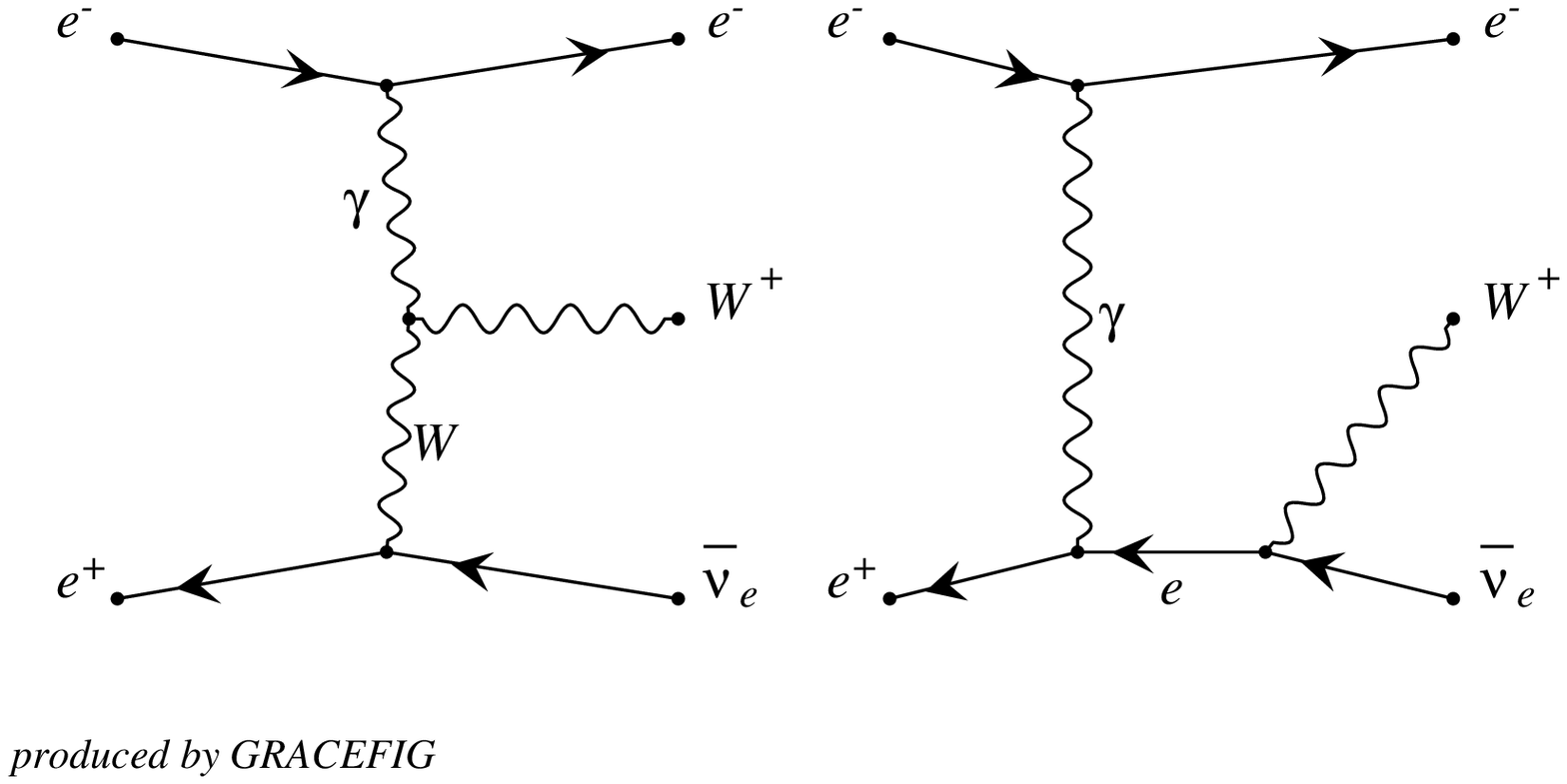,height=5.0cm}
       \vspace{-1.0cm}
      \caption{Feynman diagrams for 
        $\e^+\e^- \rightarrow \e^- \overline{\nu}_\e \W^+$.}
      \label{fig:Feynman_enw}
   \end{center}
\end{figure}

The single W production, $\e^+\e^- \rightarrow \enw$, is the standard
model process~\cite{ref:single_boson} as shown in 
Fig.\ref{fig:Feynman_enw}.   The total cross section is 
$\sigma_{\enw} = 0.6\:\pb$ at the centre-of-mass energy of $183\:\GeV$
which is much smaller than WW production 
$\sigma_{\WW} = 15.7\:\pb$.
Contributions from Z boson exchange diagrams are negligible at LEP
energies.  Thus this process offers almost pure sensitivity to the
$\WWg$ coupling~\cite{ref:Kurihara}.

\begin{figure}
   \begin{center}
        \vspace{-1.5cm}
      \epsfig{file=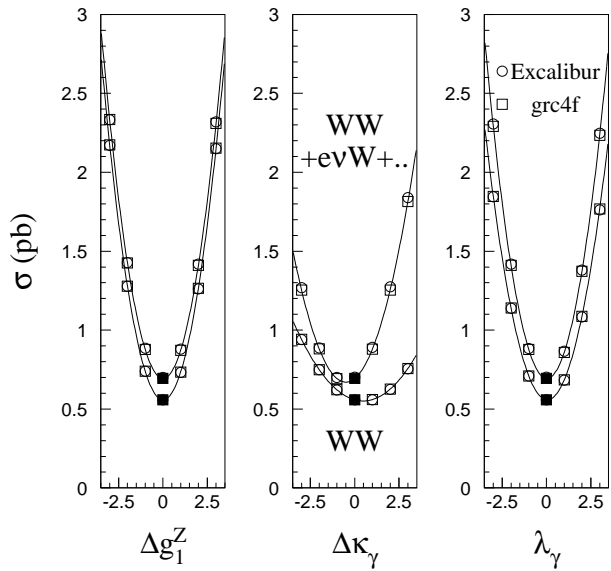,width=9cm}
      \vspace{-1.1cm}
      \caption{The total cross section for 
        $\e^+\e^- \rightarrow \e^{-} \overline{\nu}_\e \mbox{u} \bar{\mbox{d}}$
        as functions of 3 coupling parameters.   The lower curves 
        show the cross sections for W pair production alone, 
        and the upper curves are for all four fermion diagrams.   
        For one plot, the other two parameters are fixed to the
        standard model values.
        Closed points indicate the standard model prediction.}
      \label{fig:enud}
   \end{center}
\end{figure}

   The sensitivity for TGC parameters for the four fermion process of 
$\e^+\e^- \rightarrow \e^{-} \overline{\nu}_\e \mbox{u} \bar{\mbox{d}}$
is shown in Fig.~\ref{fig:enud}.   The total cross section has been
calculated with the SU(2)xU(1) constraints.   While $\WW$ production cross
section is minimum at $\dgZ = \dkg = \lg = 0$, $\enw$ production cross
section is minimum at $\dkg = -1$ and $\lg = 0$.   It can be seen that
single W production is sensitive to $\kg$, while it has the  modest
sensitivity to $\lg$.   This should be compared to the $\W\gamma$
production at hadron colliders~\cite{ref:UA2,ref:CDF,ref:D0} which is 
sensitive to $\lg$ or to 
$\mbox{b} \rightarrow \mbox{s} \gamma$~\cite{ref:CLEO,ref:ALEPH_Penguin}
which is sensitive to $\kg$ in the $\WWg$ vertex.

\begin{figure}
   \begin{center}
      \begin{tabular}{cc}
          a) $\W \rightarrow e \nu_e$ 
        & b) $\W \rightarrow \mu \nu_\mu$ \\
           \epsfig{figure=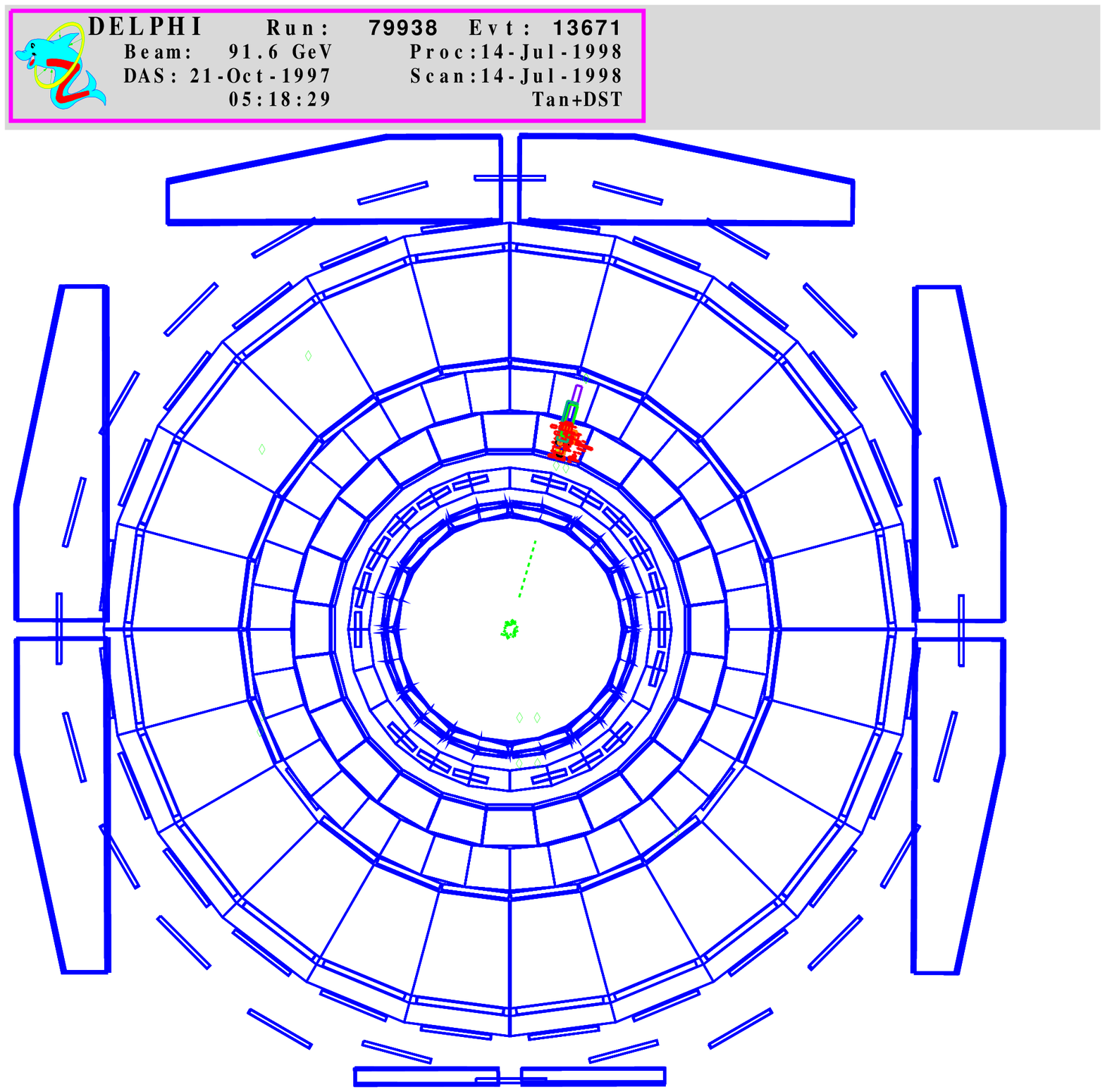,width=4.0cm}
         & \begin{sideways} \begin{sideways} \begin{sideways}
           \epsfig{figure=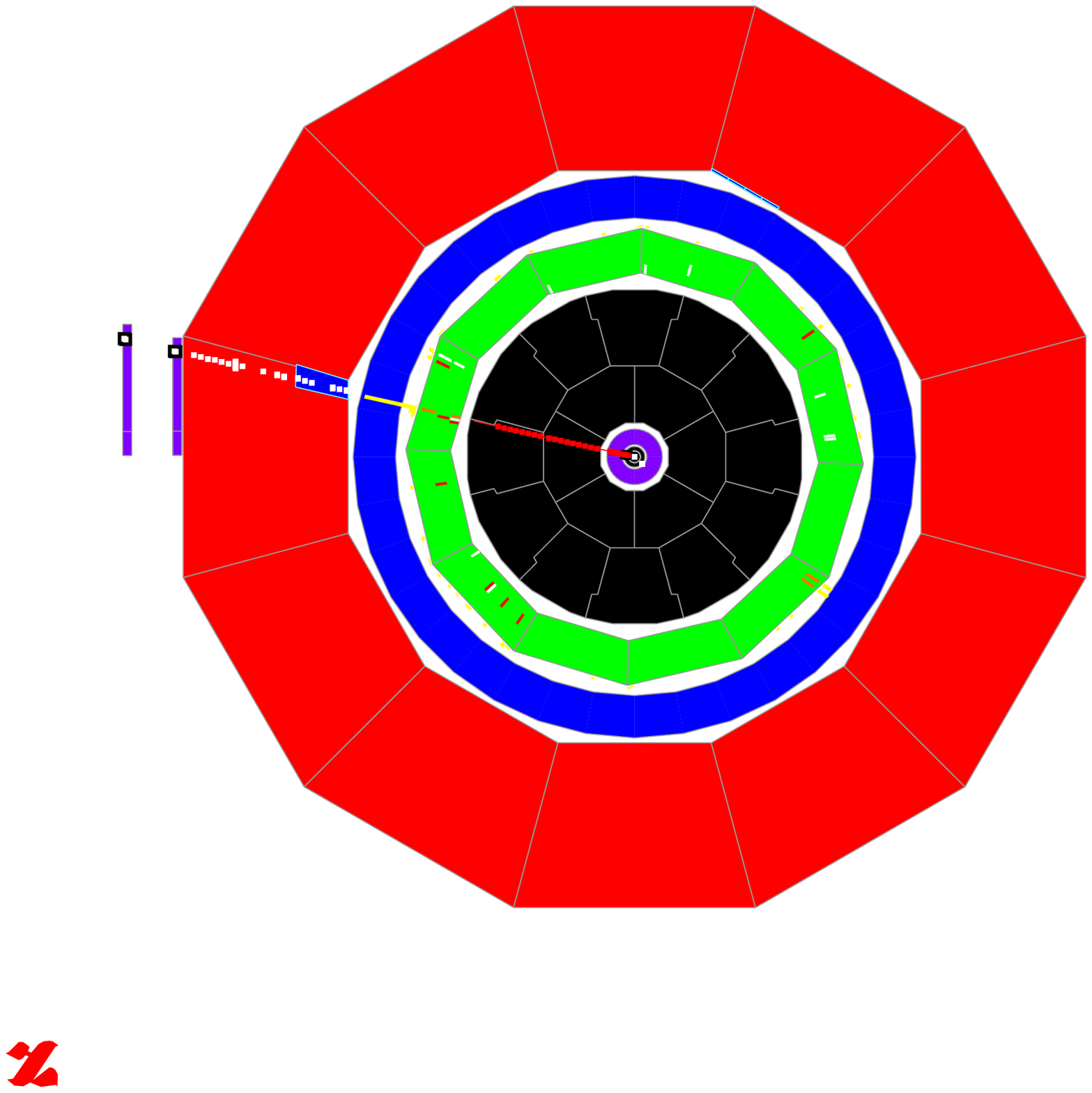,width=4.0cm}
          \end{sideways} \end{sideways} \end{sideways} \\ \\ 
          c) $\W \rightarrow \tau \nu_\tau$  
        & d) $\W \rightarrow \qq$ \\
          \epsfig{figure=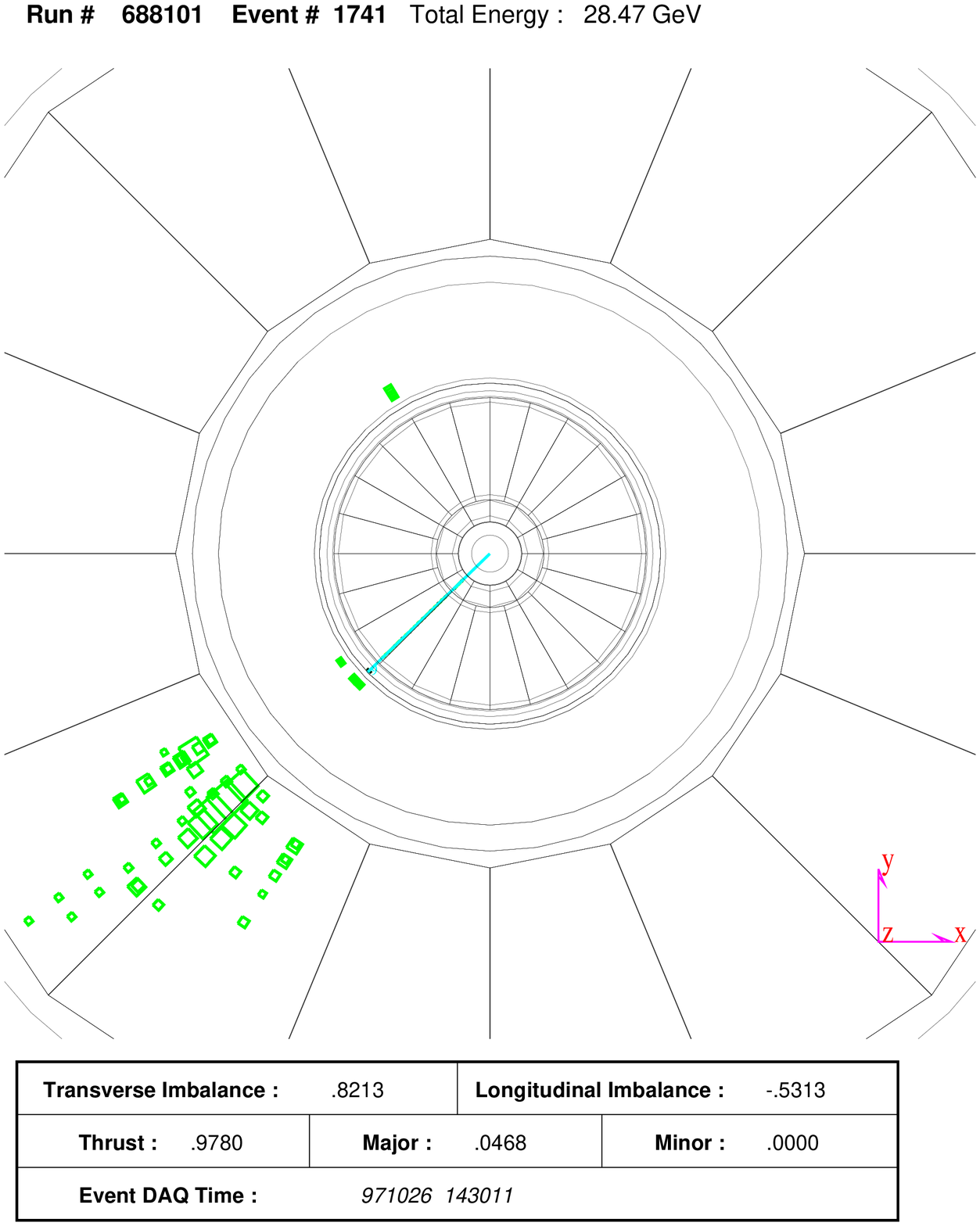,width=3.0cm} 
        & \epsfig{figure=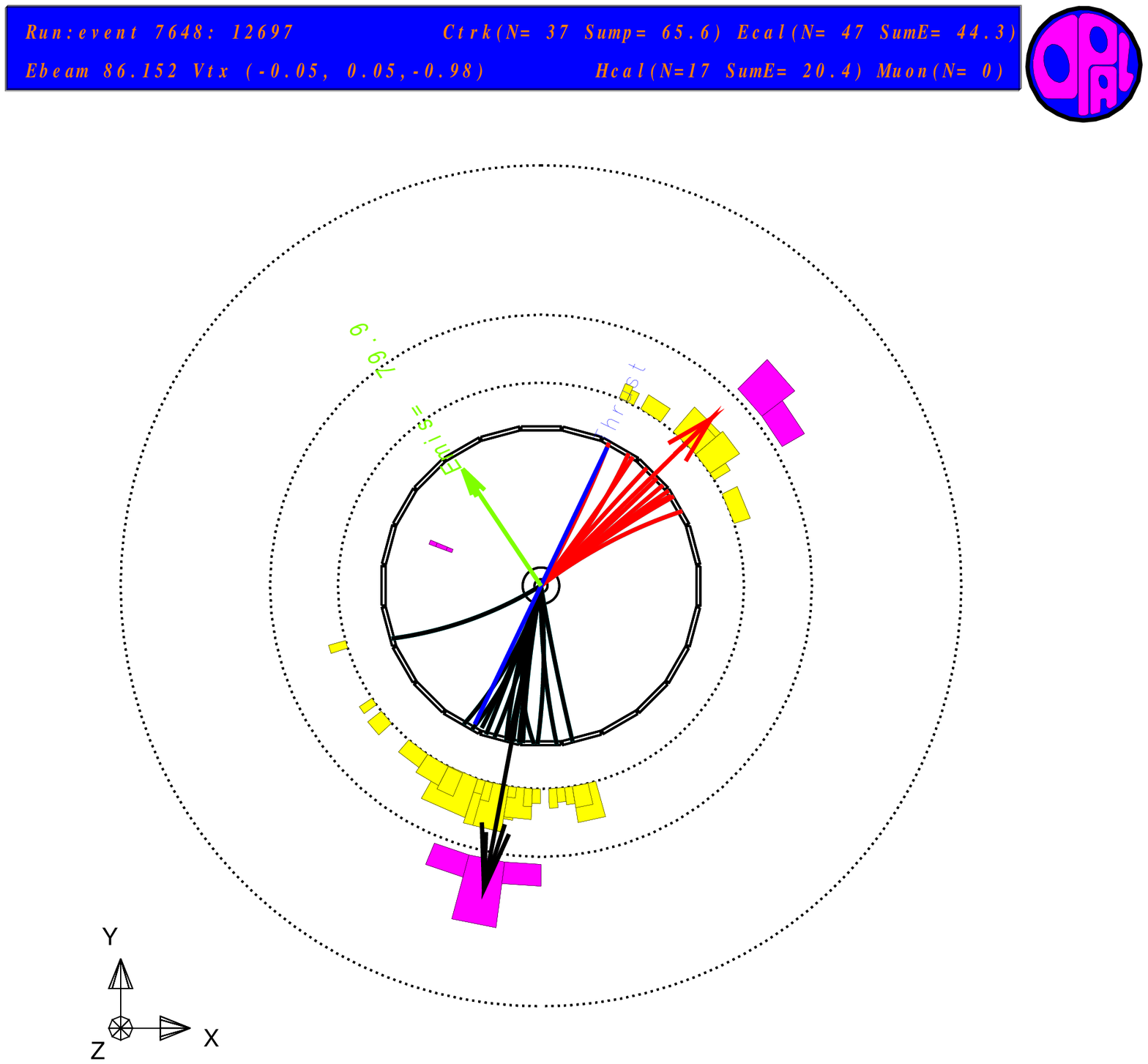,width=4.0cm} \\
      \end{tabular}     
      \vspace{0.3cm}    
      \caption{Candidate events for $\e^+\e^- \rightarrow \enw$
                observed by a)~DELPHI, b)~ALEPH, c)~L3 and d)~OPAL 
                experiments.}
      \label{fig:events}
   \end{center}
\end{figure}

\subsection{Single W signal}

Event characteristics of the single W production are as follows.
Due to its small momentum transfer, the outgoing electron escapes 
in the beam direction.   In the analysis, it is required that the
electron to be un-tagged.   This is important to suppress the
contribution from W pair production.   The associated neutrino may
carry a large transverse momentum, thus the signature of single W
production is characterized by the large missing momentum.
\noindent
For leptonic decay channel of W boson ($\W \rightarrow l \nu$),
an isolated high $P_t$ lepton with energy at about $40\:\GeV$
is the signal.   The dominant backgrounds are from $ll\nu\nu$ ($\eeZ$)
processes.
\noindent
For hadronic W decay channel ($\W \rightarrow \qq$), the signature is 
the acoplanar two jets whose invariant mass is equal to W mass.
The main background is W pair production 
($\W\W \rightarrow \tau \nu_\tau \qq $).   If $\nu_\tau$ carries away
large fraction of energy, $\tau$ becomes invisible.
It is practically impossible to distinguish this case from
the single W production, thus becoming irreducible backgrounds.

   The definitions of the single W signal are different among LEP
experiments.   The ALEPH collaboration, for example, defines the signal
as~\cite{ref:ALEPH}:
\[   \left\{
        \begin{array}{ll}
        \theta_{\mbox{e}} < 34\ \mbox{mrad}, &  \\ 
        \E_{\ell} > 20\ \mbox{GeV and } |\cos\theta_{\ell}|<0.95 
        &\ \mbox{for}\ \W \rightarrow l \nu, \\
        \M_{\mbox{qq}'} > 60\ \GeVcc &\ \mbox{for}\ \W \rightarrow \qq,
        \end{array}
   \right.      \]
where $\theta_{\mbox{e}}$ is the polar angle of the scattered electron,
$\E_{\ell}$ and $\theta_{\ell}$ are the energy and polar angle 
of leptons from the W decay, respectively. $\M_{\mbox{qq}'}$ is the 
invariant mass of the quark pair.   These cuts on W decay final states
are necessary to remove the non-resonant four fermion backgrounds.
The selected evens are displayed in Fig.~\ref{fig:events} for four
W decay modes.

Monte Carlo generators of GRC4F~\cite{ref:grc4f},
EXCALIBUR~\cite{ref:EXCALIBUR} and DELTGC~\cite{ref:DELTGC} have been
used to simulate the $\enw$ production.

\subsection{Total cross section}

\begin{figure}
   \begin{center}
       \vspace{-0.3cm}  
   \epsfig{file=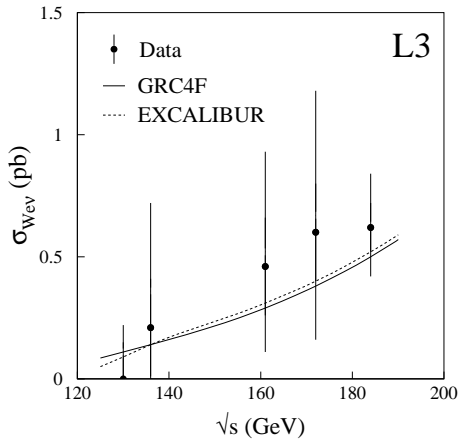,width=6.5cm}
       \vspace{-0.2cm}
   \caption{The measured cross section of $\enw$ production 
        as a function of the centre-of-mass energy by L3.}
   \label{fig:TGC_L3}
   \end{center}
\end{figure}

The summary of analyzed data and observed number of events is
given in Table~\ref{tab:smtab}.   In addition to W decay to electron
or muon, ALEPH and L3 collaborations have also analyzed the tau decay channel.
ALEPH~\cite{ref:ALEPH} has measured the total cross section of 
$\enw$ production at $183\,\GeV$ as
$\sigma_{\enw} = 0.40 \pm 0.17 (\stat) \pm 0.04 (\syst)\:\pb$
where the standard model predicts \ $0.41\:\pb$.
L3~\cite{ref:L3} has also measured the cross section at 
183\,GeV as
$\sigma_{\enw} = 0.62\ ^{+0.19}_{-0.18} (\stat) \pm 0.04 (\syst)\:\pb$
where $0.50\:\pb$ is expected from the standard model.
All these results are consistent with the standard model
expectation.
In Fig.~\ref{fig:TGC_L3}, the cross section as a function of the 
centre-of-mass energy is shown as measured by L3 experiment.   

\begin{table*}
   \begin{center}
   \caption{Summary of single W measurement for leptonic and hadronic channels.
            $\mbox{N}_{\mbox{obs}}$ is the number of selected data events.
            $\mbox{N}_{\mbox{SM}}$ and $\mbox{N}_{\enw}$
            are the expected number of total events (signal plus backgrounds)
            and $\enw$ signal events, respectively.}
   \label{tab:smtab}
   \vspace{0.4cm}
   \begin{tabular}{|l||r|r||r|rr||r|rr|} \hline
        \raisebox{0pt}[12pt][6pt]{ } & 
        \raisebox{0pt}[12pt][6pt]{$\E_{\mbox{CMS}}$} & 
        \raisebox{0pt}[12pt][6pt]{Lumi.} & 
        \multicolumn{3}{|c||}
          {\raisebox{0pt}[12pt][6pt]{$\W \rightarrow l \nu$ }} &
        \multicolumn{3}{|c|}
          {\raisebox{0pt}[12pt][6pt]{$\W \rightarrow \qq$ }} \\
        \cline{2-9}
        \raisebox{0pt}[12pt][6pt]{ } & 
        \raisebox{0pt}[12pt][6pt]{(GeV)} & 
        \raisebox{0pt}[12pt][6pt]{($\pb^{-1}$)} & 
        \raisebox{0pt}[12pt][6pt]{$\mbox{N}_{\mbox{obs}}$} &
        \raisebox{0pt}[12pt][6pt]{$\mbox{N}_{\mbox{SM}}$}  &
        \raisebox{0pt}[12pt][6pt]{$(\mbox{N}_{\enw})$}     &
        \raisebox{0pt}[12pt][6pt]{$\mbox{N}_{\mbox{obs}}$} &
        \raisebox{0pt}[12pt][6pt]{$\mbox{N}_{\mbox{SM}}$}  &
        \raisebox{0pt}[12pt][6pt]{$(\mbox{N}_{\enw})$} \\ 
        \hline  
        \raisebox{0pt}[12pt][6pt]{ALEPH~\cite{ref:ALEPH}} &
        \raisebox{0pt}[12pt][6pt]{161-183}  &
        \raisebox{0pt}[12pt][6pt]{78.9}     &
        \raisebox{0pt}[12pt][6pt]{11}       &  
        \raisebox{0pt}[12pt][6pt]{11.1}     & 
        \raisebox{0pt}[12pt][6pt]{(7.3)}    &  
        \raisebox{0pt}[12pt][6pt]{21}       & 
        \raisebox{0pt}[12pt][6pt]{21.5}     & 
        \raisebox{0pt}[12pt][6pt]{(8.8)}    \\ 
        \hline
        \raisebox{0pt}[12pt][6pt]{DELPHI~\cite{ref:DELPHI}} &
        \raisebox{0pt}[12pt][6pt]{161-183}  &
        \raisebox{0pt}[12pt][6pt]{73.0}     &
        \raisebox{0pt}[12pt][6pt]{9}        &  
        \raisebox{0pt}[12pt][6pt]{5.4}      & 
        \raisebox{0pt}[12pt][6pt]{(5.2)}    &  
        \raisebox{0pt}[12pt][6pt]{44}       & 
        \raisebox{0pt}[12pt][6pt]{52.6}     & 
        \raisebox{0pt}[12pt][6pt]{(19.9)}   \\ 
        \hline
        \raisebox{0pt}[12pt][6pt]{L3~\cite{ref:L3}} &
        \raisebox{0pt}[12pt][6pt]{130-183}  &
        \raisebox{0pt}[12pt][6pt]{88.5}     &
        \raisebox{0pt}[12pt][6pt]{12}       &  
        \raisebox{0pt}[12pt][6pt]{10.2}     & 
        \raisebox{0pt}[12pt][6pt]{(6.0)}    &  
        \raisebox{0pt}[12pt][6pt]{109}      & 
        \raisebox{0pt}[12pt][6pt]{103.3}    & 
        \raisebox{0pt}[12pt][6pt]{(14.7)}   \\ 
        \hline
        \raisebox{0pt}[12pt][6pt]{OPAL~\cite{ref:OPAL}} &
        \raisebox{0pt}[12pt][6pt]{161-172}  &
        \raisebox{0pt}[12pt][6pt]{20.3}     &
        \raisebox{0pt}[12pt][6pt]{2}        &  
        \raisebox{0pt}[12pt][6pt]{2.0}      & 
        \raisebox{0pt}[12pt][6pt]{(0.8)}    &  
        \raisebox{0pt}[12pt][6pt]{4}        & 
        \raisebox{0pt}[12pt][6pt]{2.5}      & 
        \raisebox{0pt}[12pt][6pt]{(1.3)}    \\ 
        \hline
   \end{tabular}
   \end{center}
\end{table*}

\subsection{Limits on TGC}

Since WW backgrounds in hadronic W decay channel are irreducible
(S/N=1/1 at best) and the 2/3 of W's decay hadronically, 
the pure sensitivity to $\WWg$ vertex of $\enw$ production is lost.   
This is because W pair production contains both $\WWZ$ and $\WWg$ 
vertices that cannot be separated.
One is therefore obliged to either
a) fix the irreducible WW backgrounds in hadronic W decay as
the standard model (ALEPH, OPAL),
or
b) vary the WW backgrounds simultaneously according to TGC values 
assuming SU(2)$\times$U(1) constraints (DELPHI, L3).
The former takes the conservative approach, and the latter benefits the
information contained in WW backgrounds.

\begin{figure}
   \begin{center}
       \vspace{-0.1cm}  
   \epsfig{file=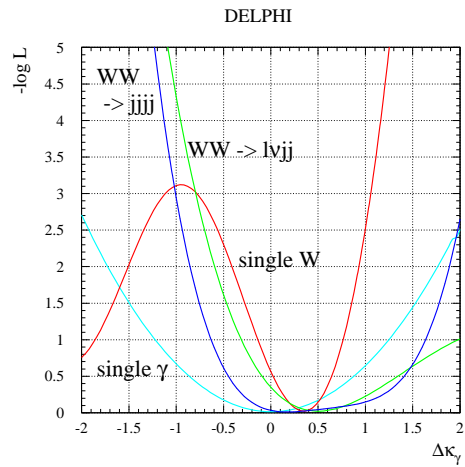,width=6cm}
   \caption{The log-likelihood functions on $\dkg$ parameter measured 
        by DELPHI.   
        The results from W pair production (hadronic, semi-leptonic), 
        single W and single gamma are shown separately.}
   \label{fig:TGC_DELPHI}
   \end{center}
\end{figure}

The sensitivity to $\kg$ of single W production which is superior to 
WW production is demonstrated in Fig.~\ref{fig:TGC_DELPHI}.   
However there are two minima at $\dkg = 0$ (the standard model) and at -2
for single W alone due to the fact that total cross section has the
same value at these points.   This double minima structure can be
solved by combining it with the results from single gamma and/or 
WW productions.

\begin{table}[b]
   \begin{center}
   \caption{The $95\%\:$C.L. limits on TGC couplings.
            Note that L3 gives $95\%\:$C.L. limits with
            2 parameter fit, and the other experiments give
            1 parameter fit result fixing the rest as 
            the standard model values.}
   \label{tab:tgc}
   \begin{tabular}{|l||c|c|} \hline
        \raisebox{0pt}[12pt][6pt]{ALEPH} & 
        \raisebox{0pt}[12pt][6pt]{$-2.6 < \dkg < 0.5$} & 
        \raisebox{0pt}[12pt][6pt]{$-1.6 <  \lg < 1.6$} \\ 
        \hline  
        \raisebox{0pt}[12pt][6pt]{DELPHI} & 
        \raisebox{0pt}[12pt][6pt]{$-0.4 < \dkg < 0.9$} & 
        \raisebox{0pt}[12pt][6pt]{$-1.5 <  \lg < 1.5$} \\ 
        \hline  
        \raisebox{0pt}[12pt][6pt]{L3} & 
        \raisebox{0pt}[12pt][6pt]{$-0.46 < \dkg < 0.57$} & 
        \raisebox{0pt}[12pt][6pt]{$-0.86 < \lg  < 0.75$} \\ 
        \hline  
        \raisebox{0pt}[12pt][6pt]{OPAL} & 
        \raisebox{0pt}[12pt][6pt]{$-3.6 < \dkg < 1.6$} & 
        \raisebox{0pt}[12pt][6pt]{$-3.1 <  \lg < 3.1$} \\ 
        \hline  
   \end{tabular}
   \end{center}
\end{table}

In Table~\ref{tab:tgc}, the limits on TGC parameters are summarized.
The event yields have been analyzed by Bayesian approach (ALEPH) or by
maximum likelihood fits to event rate (OPAL) and to kinematical
distributions (DELPHI, L3).   
One should note that the results on $\lg$ obtained by DELPHI and L3 
experiments benefit information contained in W pair production.
The intrinsic sensitivity of single W production alone at 
the current statistics of LEP is $|\dkg| < 0.5$ ($\lg = 0$) and 
$|\lg| < 1.6$ ($\dkg = 0$) at $95\%\:$C.L.
provided that the double minima structure for $\dkg$ is resolved.

\section{$\nng$ Production}

\begin{figure}[h]
    \begin{center}      
      \vspace{-0.5cm}   
      \epsfig{file=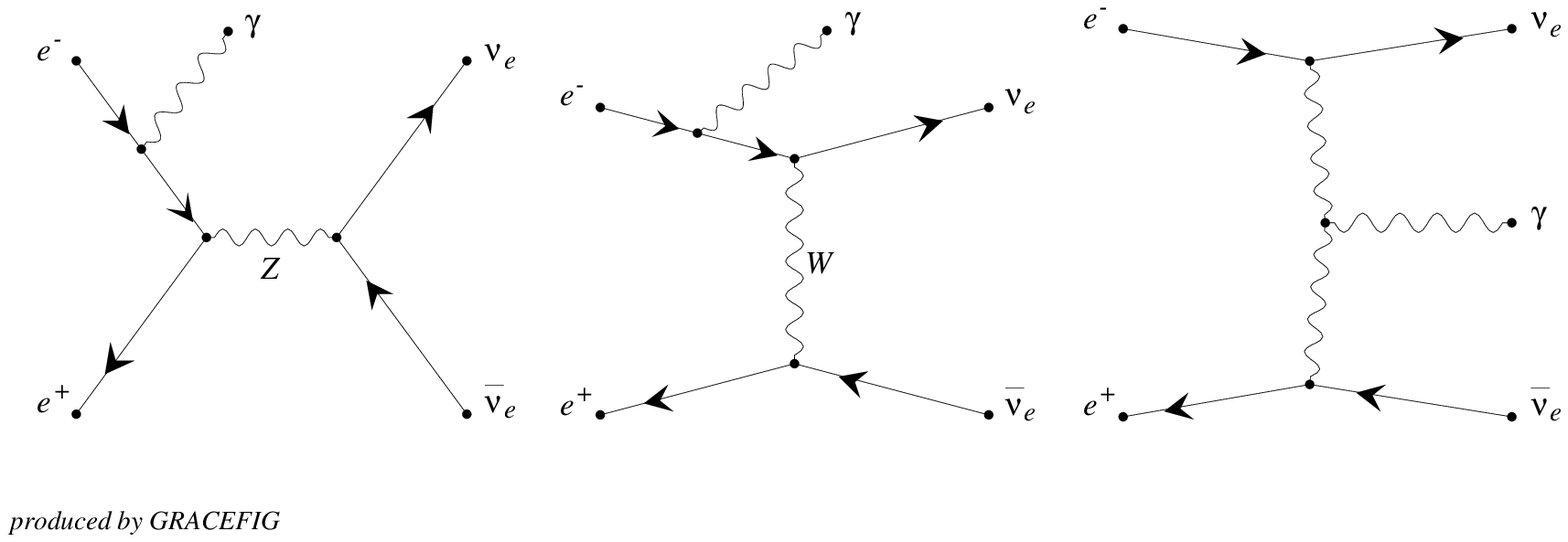,width=9cm,height=4cm}    \\
      \vspace{-2.85cm} \hspace{4.7cm} {\tiny\bf $W$} \\
      \vspace{0.5cm}   \hspace{4.7cm} {\tiny\bf $W$} \\
      \vspace{1.0cm}    
      \caption{Feynman diagrams for 
        $\e^+\e^- \rightarrow \nu_e \bar{\nu}_e \gamma$.}
      \label{fig:Feynman_nng}
   \end{center}
\end{figure}

Amongst various physics opportunities such as counting the number of
light neutrino species, the process $\e^+\e^- \rightarrow \nng$ is
also sensitive to $\WWg$ coupling~\cite{ref:single_gamma}.
There are three types of diagrams which contribute to the $\nng$ final
state as shown in Fig.~\ref{fig:Feynman_nng}.
The first diagram is the radiative return to Z by emitting hard
photon, the second one is t-ch W boson exchange, and the last one
is W boson fusion type which contains a $\WWg$ vertex.
The photon in the radiative return process has energy peaked at 
$\mbox{x}_\gamma = \E_\gamma/\E_{\mbox{beam}} = 0.74\ \mbox{at}\ 183\:\GeV$.
Monte Carlo programs based on KORALZ~\cite{ref:KoralZ} and
DELTGC~\cite{ref:DELTGC} are used.

\begin{figure}
   \begin{center}
       \vspace{-1.2cm}
    \epsfig{figure=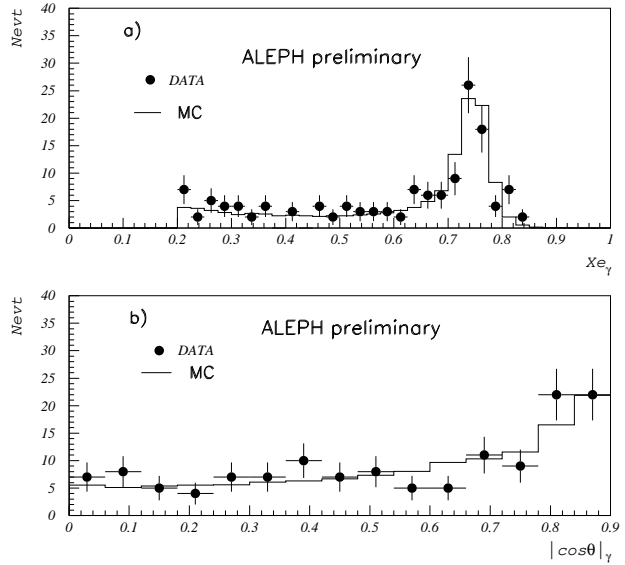,width=9cm}
       \vspace{-0.7cm}
   \caption{ALEPH measurements on 
        a)~photon energy (normalized to the beam energy)
        and b)~angular distribution of photons for $\nng$ production
        at $183\,\GeV$.}
   \label{fig:NNG_ALEPH}
   \end{center}
\end{figure}

   Isolated photons have been searched in the analysis.   It is found
that the data are in good agreement with the standard model expectation.
\noindent
When extracting coupling parameters with the maximum likelihood method, 
the total yield of observed events, the energy and angular distributions 
are used as shown in Fig.~\ref{fig:NNG_ALEPH}.   The photon energy
region of $\mbox{x}_\gamma \in [0.67,0.76]$ is not used in ALEPH's analysis. 
ALEPH~\cite{ref:ALEPH_gam} has obtained the fitted results of
$\dkg =  0.05^{+1.2}_{-1.1} \pm 0.3\ (\lg = 0)$ and
$\lg  = -0.05^{+1.6}_{-1.5} \pm 0.3\ (\dkg = 0)$, 
where the first error is statistical and the second is systematic.
\noindent
DELPHI~\cite{ref:DELPHI} performs the binned likelihood fit to the
whole photon energy spectrum, and gets
$\dkg =  0.00^{+1.01}_{-1.01} \pm 0.36\ (\lg = 0)$ and
$\lg  =  0.72^{+1.12}_{-1.12} \pm 0.36\ (\dkg = 0)$.
Both results are consistent with the coupling parameters equal to zero.

The sensitivity of $\nng$ to TGC parameters is $2 \sim 3$ times weaker
than that of $\enw$, but nevertheless, it contributes to solve the
double minima structure for $\dkg$ in $\enw$ production as discussed above.

\section{Invisible W Decay}

The ALEPH collaboration~\cite{ref:ALEPH}
has performed the search for the invisible W decay
in $\e^+\e^- \rightarrow \W^+ \W^-$.   The mixed supersymmetric/standard 
model decay has been studied.    One W boson decays to chargino and
neutralino, followed by the chargino decay to sneutrino and lepton.
The other W boson decays to the standard model particles.
The whole decay cascade can be illustrated as,
\[ \e^+\ \e^- \rightarrow 
\begin{array}[t]{ll}
   \W            & \W \\
   \hookrightarrow \mbox{\large SM}  & \hookrightarrow 
                  \begin{array}[t]{rl}
                     \chi^\pm \ \chi & \\
                     \hookrightarrow \ell & 
                     \begin{array}[t]{ll}
                        \widetilde{\nu} & \\
                        \hookrightarrow \nu\ \chi. &
                     \end{array} 
                  \end{array}
\end{array} \]
The supersymmetric decay of W boson becomes practically invisible if
the mass difference between the chargino and the sneutrino 
($\dM \equiv m_{\chi^\pm} - m_{\widetilde{\nu}}$) is less than about 
$3\,\GeVcc$.   However this process still can be tagged by the  
$\W$ decay to the standard model particles.   Three event topologies of
the final state, single lepton ($e/\mu$), acoplanar lepton pair 
(one is soft) and hadrons (missing mass equal to $\W$ mass) have been 
studied.

No excess of the signal has been observed and the
results are consistent with the standard model expectation.
The limits at 95\% C.L. on the W boson supersymmetric branching ratio
have been obtained as:
\[ \begin{array}{llcl}
   {\cal B}_{susy} & (\dM \approx 0\,\GeVcc) & < & 1.3\,\%,\\[5pt]
   {\cal B}_{susy} & (\dM = 3\,\GeVcc) & < & 1.9\,\%,\\
\end{array} \]
\noindent
assuming ${\cal B}(\chi^\pm\rightarrow\ell\tilde{\nu})=100\%$ and
$m_{\chi^{\pm}} = 45\,\GeVcc$.
Degenerate ($\dM \approx 0\,\GeVcc$) case gives the
quasi model-independent limit on the invisible W decay width via
direct search.   The result translates as 
$\Gamma(\W \rightarrow \mbox{inv}) < 27\,\MeV$ at 95\% C.L..

\section{Conclusion}

   Single W production has been studied at LEP.   The production cross
section is consistent with the standard model expectation.   
It has been shown that $\enw$ production is sensitive to the $\WWg$
coupling, in particular to $\kg$.   However, the irreducible WW
background in hadronic decay channel of $\enw$ does not allow the
clear separation of $\WWg$ and $\WWZ$ couplings.
Single gamma production has also been studied.    No deviation from
the standard model is found.

   Search for invisible W decay has been performed by ALEPH 
and the stringent limit on invisible W decay width of $27\:\MeV$
has been obtained at $95\,\%\:$C.L..

The current status and the future perspective on the $\WWg$ coupling
measurement are summarized in Table~\ref{tab:future}.
One sees that the $\W\gamma$ production at Tevatron and 
$\enw$ production at LEP provide complementary information on TGC.
It is anticipated that $\enw$ production at LEP has the sensitivity of
$|\dkg| \sim 0.1$ with $500\:\pb^{-1}$ data at higher energies. 
In future, one may combine leptonic decay channel of $\enw$ and $\nng$
alone that are purely sensitive to the $\WWg$ coupling.    
It is expected that the use of kinematical information and the spin 
analysis will further improve the limits.

\begin{table}[h]
   \begin{center}
   \vspace{-0.3cm}
   \caption{The current and future TGC limits at $95\%\:$C.L.
            par single experiment.}
   \label{tab:future}
   \vspace{-0.1cm}
   \begin{tabular}{|l||l|r||l|l|} \hline
        \raisebox{0pt}[12pt][6pt]{Tevatron~\cite{ref:D0}}      & 
        \raisebox{0pt}[12pt][6pt]{$\W\gamma$}                  & 
        \raisebox{0pt}[12pt][6pt]{$93\:\pb^{-1}$}              & 
        \raisebox{0pt}[12pt][6pt]{$\left| \dkg \right| < 0.9$} & 
        \raisebox{0pt}[12pt][6pt]{$\left| \lg  \right| < 0.3$} \\ 
        \hline  
        \raisebox{0pt}[12pt][6pt]{LEP}                         & 
        \raisebox{0pt}[12pt][6pt]{$\enw$}                      & 
        \raisebox{0pt}[12pt][6pt]{$80\:\pb^{-1}$}              & 
        \raisebox{0pt}[12pt][6pt]{$\left| \dkg \right| < 0.5$} &
        \raisebox{0pt}[12pt][6pt]{$\left| \lg  \right| < 1.6$} \\
        \hline \hline
        \raisebox{0pt}[12pt][6pt]{LEP2000}                     & 
        \raisebox{0pt}[12pt][6pt]{$\enw$}                      & 
        \raisebox{0pt}[12pt][6pt]{$500\:\pb^{-1}$}             & 
        \raisebox{0pt}[12pt][6pt]{$\left| \dkg \right| < 0.1$} &
        \raisebox{0pt}[12pt][6pt]{$\left| \lg  \right| < 0.6$} \\
        \hline
   \end{tabular}
   \end{center}
\end{table}

\vspace{-0.5cm}
\section*{Acknowledgements}

   The author would like to thank following single W'ers and single
$\gamma$'ers at LEP for discussions and correspondences, 
J.~Boucrot, J.B.~Hansen, A.~Jacholkowska and D.~Zerwas (ALEPH), 
C.~Matteuzzi, R.L.~Sekulin and O.~Yushchenko (DELPHI),
P.~de Jong and A.~Kounine (L3), 
G.~Bella and M.~Verzocchi (OPAL).



\end{document}